# Two magnetic regimes in doped ZnO corresponding to a dilute magnetic semiconductor and a dilute magnetic insulator

A.J. Behan, A. Mokhtari, H.J. Blythe, D. Score, X-H. Xu,[1] J.R. Neal, A.M. Fox and G.A. Gehring[2]

Department of Physics and Astronomy, The University of Sheffield, Sheffield, S3 7RH, UK.
[1]Permanent address: School of Chemistry and Materials Science, Shanxi Normal University, Linfen, Shanxi 041004, PR China.

**Abstract**

Films of ZnO doped with magnetic ions, Mn and Co and, in some cases, with Al have been fabricated with a very wide range of carrier densities. Ferromagnetic behaviour is observed in both insulating and metallic films, but not when the carrier density is intermediate. Insulating films exhibit variable range hopping at low temperatures and are ferromagnetic at room temperature due to the interaction of the localised spins with static localised states. The magnetism is quenched when carriers in the localised states become mobile. In the metallic (degenerate semiconductor) range, robust ferromagnetism reappears together with very strong magneto-optic signals and room temperature anomalous Hall data. This demonstrates the polarisation of the conduction bands and indicates that, when ZnO is doped into the metallic regime, it behaves as a genuine magnetic semiconductor.

[2]) Corresponding author: FAX (44) 114 222 3555; electronic address g.gehring@shef.ac.uk

The search for spintronic materials that combine both semiconducting and ferromagnetic properties is currently one of the most active research fields in magnetism. Compounds based on ZnO are especially exciting in this context since, in contrast to GaMnAs and InMnAs, they exhibit ferromagnetism at room temperature [1-7]. Despite the progress in developing ZnO as a spintronic material, there has been much controversy concerning the mechanism that causes the magnetism [8-10]. It has been found that not all doped films exhibit ferromagnetism, and that the mobile carrier density, $n_c$, can be very different in those compounds that do. This implies that the established theory of carrier-mediated magnetism, which works well for p-type GaMnAs, is not generally applicable to this n-type material. For example, it has been found that the addition of Zn interstitials, which affects both the number of neutral and ionised donors, leads to an increase in the ferromagnetism [11], and a recent study of Al-doped ZnCoO has reported a variation of magnetisation with Al content rather than carrier density [12]. In fact, most of the work on doped ZnO has concentrated on the insulating phase [9,13-15], to such an extent that some authors now refer to ZnO as a dilute magnetic insulator (DMI) [13]. However, we have recently reported the observation of ferromagnetism in Al-doped films where $n_c$ is very high [16], which highlights the importance of exploring the full range of carrier densities from the insulating to the metallic phases.

In this Letter we report a systematic study of the relationship between the magnetism and conductivity in transition metal (TM) doped ZnO. By studying a large number of films with a wide range of carrier densities, we have identified three distinct conductivity regimes:
1. The *insulating* phase at low carrier densities, in which the conductivity at low temperatures arises from a variable range hopping (VRH) process [17]. In this regime, labelled VRH, the least conducting films are the most magnetic, as has been observed previously [13,14].
2. The *intermediate* regime, labelled I, where the samples satisfy neither the conditions for VRH at low temperatures nor for metallic behaviour at room temperature. Here, we find that the magnetism decreases and finally disappears as the films are doped out of the insulating state, in agreement with previous reports [18,19], and then reappears as the films become metallic.
3. The *metallic* phase (labelled M) at high carrier densities, in which we find that the magnetism depends on the ratio of the density of free carriers to magnetic ions, as predicted by the theory of carrier-mediated exchange [20]. The large magnetic circular dichroism (MCD) signal observed at the band edge demonstrates the strong spin splitting of the ZnO conduction band.

The systematisation into three conductivity regimes explains, to a large extent, the great divergence of data that is reported in the literature.

Doped ZnO samples were grown as thin films on c-cut sapphire substrates at 450$^0$C by pulsed laser deposition (PLD) using a XeCl laser at 308nm. The oxygen pressure was varied from $10^{-5}$ to $10^{-1}$ Torr to control the oxygen stoichiometry. Co and Mn were used as the dopants with concentrations up to 6% and also with varying concentrations of Al. (All quoted concentrations refer to the targets). Magnetization measurements were made with a SQUID magnetometer and the thickness of the films was measured with a Dektak profilometer and lay in the range 100−500 nm. Hall effect measurements were made using the van der Pauw four-probe configuration in a continuous-flow He cryostat in the range 5 – 300 K with fields up to 1 T. The magneto-optic spectra were taken with a Xe lamp and monochromator with a photoelastic modulator.

The key result of our work is presented in Fig. 1, which shows the magnetization as a function of the carrier density, $n_c$, for Co-doped ZnO. The changes in carrier density occurred as a result of changing the oxygen pressure and/or the Al concentration [16]. We find that the magnetization is large at the two extremes, but vanishes at intermediate densities. As with other authors [9], we found that the reproducibility of the films was not good, and that both the magnetization and the carrier density varied from run to run. However, when the magnetization is plotted against the carrier density, a clear pattern emerges. We stress that Fig. 1 includes the results for *all* films investigated.

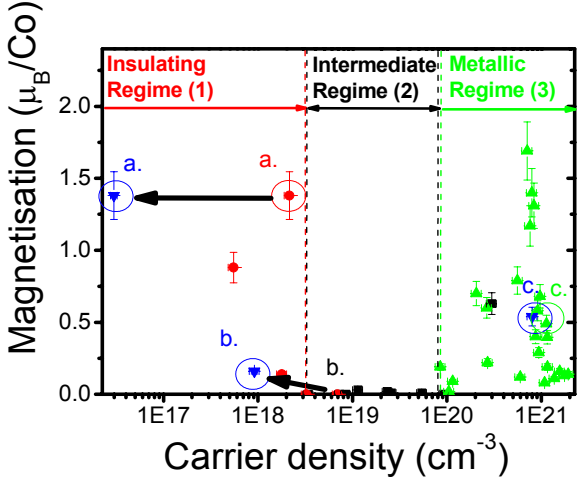

**Figure 1.** (in colour online) Room temperature magnetization for ZnO films with 5% Co and varying Al doping, as a function of the mobile carrier density $n_c$. Samples are classified by their resistivities as M for metallic, I for intermediate and VRH for variable range hopping. The metallic samples, labelled ▲, have a mean free path greater than 0.3 nm at 295 K. The VRH samples, labelled ●, followed eqn (1) over a range $5\ K < T < T^*$, where $T^* < T_0$. The intermediate samples, labelled ■, satisfied neither criterion. Magnetisation and $n_c$ data points taken at 5K for three representative films, **a**, **b** and **c** are labelled ▼. Arrows lead from these points to their equivalent data at 295 K.

The VRH regime was identified by measuring the resistivity, $\rho$, as a function of the temperature, $T$, and comparing to the Mott formula in the low temperature limit,

$$\rho(T) = A\exp(T_0/T)^{1/4}, \quad (1)$$

where $A$ and $T_0$ are fitting parameters [21] and the fit is good over a range $0 < T < T^*$ where $T^* \ll T_0$ (see insert to Fig. 2). In this regime, films generally had a room temperature resistivity above 0.1 Ω cm that increased strongly at lower temperature.

The metallic regime was identified by two tests: (i) the Fermi temperature should be high compared to room temperature, and (ii) the mean free path at room temperature, $\lambda$, estimated using the Drude formula, $\lambda = \dfrac{\hbar k_F}{ne^2\rho}$ (where $n$ and $\rho$ are the measured carrier density and resistivity and $k_F$ is found from $n$), should be greater than the lattice spacing (3Å) [17]. All the samples that satisfied these two criteria contained Al and had carrier densities above $8\times10^{19}\ cm^{-3}$. The resistivity of these samples was typically below $10^{-4}$ Ω cm and almost all had $d\rho/dT > 0$ for $5\ K < T < 300\ K$. The remaining films, where the estimated mean free path at room temperature was <3Å or the resistance was rising at low temperature but failed the VRH test ($T^* < T_0$), were classed as intermediate.

We begin by discussing the insulating regime. Here, the electron density is controlled by the dopants, the oxygen stoichiometry and other defects. The Fermi level is located below the impurity-band mobility edge and electrons are localised, thus giving rise to VRH conduction at low temperatures. The inset to Fig.2 shows a plot of $\ln(\rho)$ versus $T^{-1/4}$ for two typical films. The straight-line fit establishes the VRH behavior for $T<T^*$, with the fitting parameter, $T_0$, giving an indication of the hopping probability [21]. Fig. 2 is a plot of magnetization versus $T_0$, in samples with different dopants. We find that the samples are only strongly magnetic at room temperature if $T_0 >$10,000 K, and that, once the threshold has been passed, the magnetization increases with $T_0$ for all dopants. Since a high value of $T_0$ corresponds to a low probability for hopping and to a high resistance, our results are in agreement with refs [18] and [14], where it was observed that the magnetization and the resistivity fall when carriers are added by doping with Sn and annealing in a reducing atmosphere respectively.

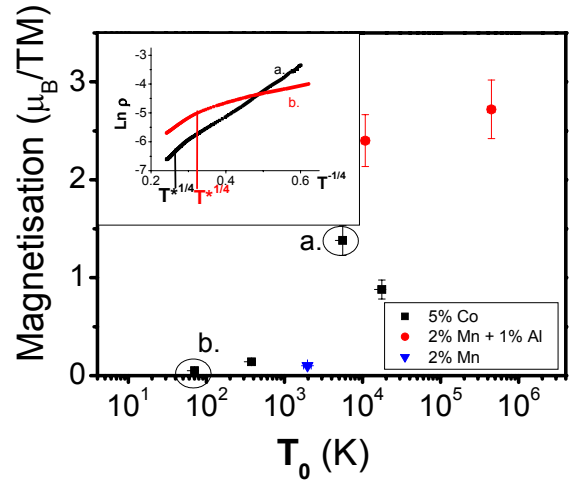

**Figure 2**. (in colour online) Room temperature magnetization for different dopants as a function of $T_0$. The inset shows $\ln(r)$ versus $T^{-1/4}$ for the two $Zn_{0.95}Co_{0.05}O$ films labelled **a** and **b** in Fig. 1. Note that film **a,** which has a much larger $T_0$, shows ferromagnetism at 295 K but film **b** does not.

The link between magnetism and $T_0$ can be understood by extension of the static magnetic polaron theory of Coey *et al.* [5]. In this model, the exchange is mediated by occupied polaron orbits, and ferromagnetism occurs when the polaronic density reaches the percolation limit. If the polaronic density changes suddenly near a magnetic ion due to the hopping of a mobile carrier, the time-dependent exchange field may flip the spin of the magnetic ion and hence destroy the magnetism. This is likely to be most effective when the carrier dwell time is comparable to the inverse of the Larmor frequency of the magnetic ions precessing in the exchange field, as occurs in the intermediate phase [22]. Thus immobile carriers can contribute to the exchange, whereas mobile carriers are inimical to magnetism.

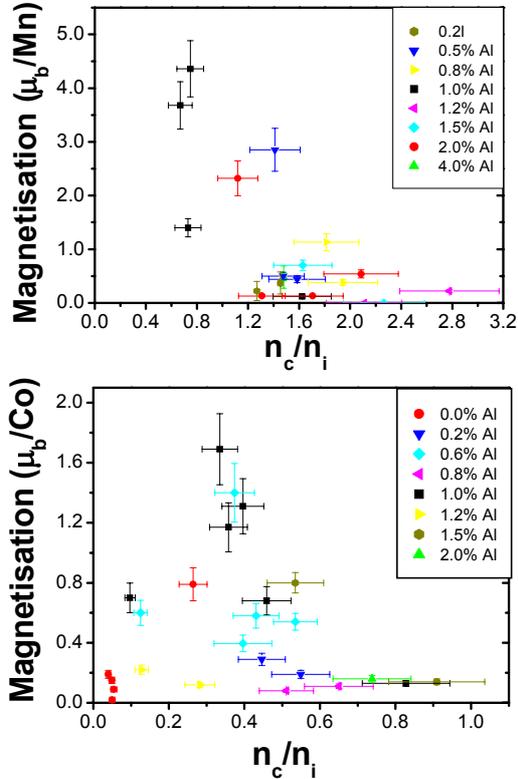

**Figure 3.** (in colour online) Room temperature magnetization as a function of the ratio of carrier density to the number of TM ions at 295 K for (a) metallic samples with 2% Mn, and (b) metallic samples with 5% Co. The legend gives the Al concentration in each case – the different carrier densities for the same Al doping resulted from changing the pressure of oxygen in the growth chamber.

Now let us consider the metallic regime. The localised spins of the TM ions are interacting with band electrons, and the standard theory for a DMS should apply [20]. There should therefore be no time dependent effects from the extended states. Theory predicts that the exchange is maximized when the ratio of $n_c$ to $n_i$, the density of magnetic ions, is between about 0.3 and 0.5. [19]. This prediction is borne out by the data presented in Fig. 3, which shows substantial magnetic moments for *all* of our Co- and Mn-doped films that are in the metallic regime. The decrease of the magnetism with increasing carrier density for $n_c / n_i > 0.5$ or 0.8 for Co- or Mn-doping respectively is apparent [16, 20]. There is considerable scatter and thus the challenge to grow reproducible high moment films remains.

We demonstrate the polarization of the carriers in the metallic phase by measuring the magnetic circular dichroism (MCD) and anomalous Hall effect (AHE). MCD data on $Zn_{0.972}Mn_{0.02}Al_{0.008}O$ and $Zn_{0.944}Co_{0.05}Al_{0.006}$ films taken with $B = \pm 0.5T$ are presented in Fig 4; these measurements made at room temperature are similar to the results obtained previously at 10K [7]. The large MCD signal seen at the ZnO band edge (3.35eV) is a clear indication that the ZnO band electrons are polarised and hence that the magnetisation is intrinsic to doped ZnO and not due to defect phases.

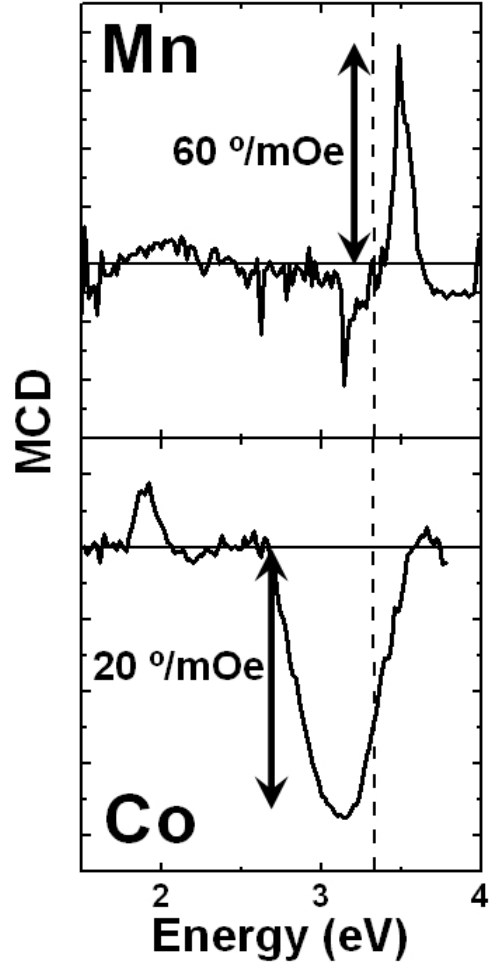

**Figure.4.** Room temperature magnetic circular dichroism (MCD) spectra for a $Zn_{0.972}Mn_{0.02}Al_{0.008}O$ and $Zn_{0.944}Co_{0.05}Al_{0.006}$ films grown at 10mTorr. The room temperature magnetisations were 1.4 $\mu_B$/Mn and 0.40$\mu_B$ /Co respectively. The vertical dashed line indicates the ZnO band edge (3.35eV).

We can eliminate metallic Co clusters as the primary source of magnetism in our Co doped samples [12] by considering the $Co^{2+}$ d-d* feature at 1.95eV The strength of this feature at room temperature is larger than expected for paramagnetic $Co^{2+}$ ions [23], thus indicating that the $Co^{2+}$ ions experience a strong exchange field from the feromagnetism. When the temperature is reduced to 10K the magnitude of this MCD feature increases by a factor ~ 4 not 30 as would occur if the Co ions were paramagnetic. The AHE measurements on ferromagnetic $Zn_{0.975}Mn_{0.02}Al_{0.005}O$ and $Zn_{0.944}Co_{0.05}Al_{0.006}O$ shown in Fig.5 give a value of 0.02 T for the coercive fields which agreed with that measured by the SQUID. No such signal was seen for paramagnetic samples. The small magnitude of the AHE in our samples is typical of oxide magnets [24,25].

Finally we consider the effect of lowering the temperature to 5K for three representative samples labelled *a*, *b* and *c* in Fig. 1. Sample *b* is non-magnetic with an intermediate carrier density at room temperature; lowering the temperature reduces its carrier density so that it becomes magnetic with a moment 0.13 $\mu_B$/Co at 5 K. Sample *a* is magnetic and insulating at room temperature, so although cooling to 5K reduces the carrier concentration, it does not increase the magnetisation any further. Sample *c* is strongly metallic and neither the carrier concentration nor the magnetism change as it is cooled.

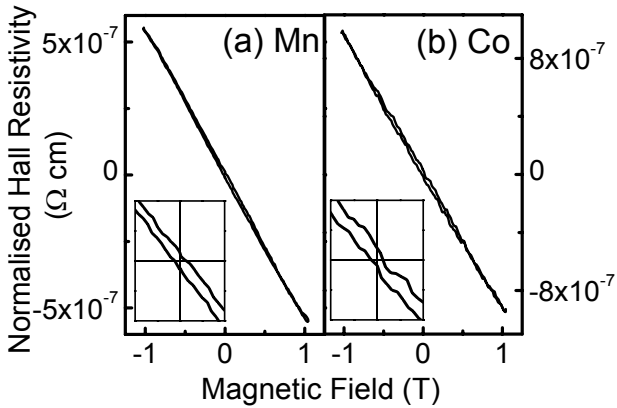

**Figure.5.** (in colour online) Room temperature normalised Hall resistivity for ferromagnetic samples showing the AHE: (a) 2% Mn + 0.5% Al, magnetisation 2.85 $\mu_B$/Mn and (b) 5% Co + 0.6% Al, magnetisation 0.58 $\mu_B$/Co. The inset data, which expands the region between ± 0.14 T, clearly shows an open loop about the origin which closes at higher fields.

It is instructive to compare our results with those of Kittilstved *et al.* [11]. In their work, the conductivity of $Co^{2+}$:ZnO was controlled by varying the density of Zn interstitials, and both the magnetism and conductivity were positively correlated with the neutral Zn donor density. In our system, the changing factor is the fraction of ionized donors that initially has a detrimental affect on the magnetism and is only neutralized when the predominant interaction of the TM ions is with extended states. It is important to note that both the carrier density and the conductivity in our metallic phase are ~100 times higher than in ref. [11], and the magnetization was also higher. Thus it appears that there are different mechanisms at work when the carriers are produced by neutral donors (i.e. Zn) and ionized donors (e.g. Al).

In conclusion, we have shown that there are two distinct mechanisms that can give rise to ferromagnetism in doped ZnO: magnetic polarons and carrier-mediated exchange. These two mechanisms give rise to DMI and DMS behaviour, respectively. In the DMI limit, the carrier density is low, the conduction at low temperatures is described by VRH theory, and the magnetization at room temperature rises with the hopping parameter $T_0$. These results can be explained by distinguishing between the statically-occupied polaron states and those that are contributing to the hopping conductivity. At intermediate carrier densities, the samples are non-magnetic. Finally, at high carrier densities, the magnetization depends on the carrier density, in agreement with the standard theory of DMS. The categorization of the samples helps to explain the very large scatter of results reported in the literature. Furthermore, the clear demonstration of carrier-mediated ferromagnetism at high carrier densities implies that heavily-doped ZnO can be used as an *n*-type DMS at room temperature, with consequent potential for exploitation in spintronic applications.

**Acknowledgments:**
We acknowledge support from the EPSRC via grants EP/D070406/1 and EP/D037581/1, also the Royal Society for a visiting fellowship for X-H Xu. We would like to thank T.M. Searle for helpful discussions.

**References**

[1] P. Sharma et al., *Nat. Mater.* **2**, 673 (2003)
[2] M. Venkatesan et al, *Phys. Rev. Lett.* **93**, 177206 (2004)
[3] H.J. Blythe et al, *J. Magn. Magn. Mater.* **283**, 117 (2004)
[4] Ü. Özgür et al, *J. Appl. Phys.* **98**, 041301 (2005)
[5] J.M.D. Coey et al, *Nat. Mater.* **4**, 173 (2005)
[6] N.S. Norberg et al, *J Am Chem Soc*, **126**, 9387 (2004)
[7] J.R. Neal et al, *Phys. Rev. Lett.* **96**, 197208 (2006)
[8] S.A. Chambers, *Surface Science Reports* **61**, 345 (2006)
[9] J.M.D. Coey, *Current opinion in Solid State and Mat. Sci.* **10**, 83 (2006)
[10] R. Seshadri, *Current opinion in Solid State and Mat. Sci.* **9**, 1 (2005)
[11] K. K. Kittilstved et al, *Phys. Rev. Lett.* **97**, 037203 (2006)
[12] M. Venkatesan et al, *Appl. Phys. Lett.* **90**, 242508 (2007)
[13] C Song et al *Phys. Rev. B*, **73**, 024405 (2006)
[14] N. Khare et al, *Adv. Mat.* **18**, 1449 (2006)
[15] D. McManus et a*l*, *Adv. Mater.*, in press
[16] X.H. Xu et al, *New J. Phys.* **8**, 135 (2006)
[17] N.F. Mott, *Metal Insulator Transitions*, 2nd edition (Taylor and Francis London , 1990)
[18] M. Ivill et al, *J. Appl. Phys.* **97**, 053904 (2005)
[19] Q. Xu et a*l*, *Phys. Rev. B* **73**, 205342 (2006)
[20] A. Chattopadhyay et al, *Phys. Rev. Lett.* **87**, 227202 (2001)
[21] S-s. Yan et al, *J. Phys.: Cond. Mat.* **18**, 10469 (2006)
[22] G.A. Gehring (to be published)
[23] W. Pacuski et al *Phys Rev* **B73**, 035214 (2006)
[24] Q. Xu et al, *J. Appl. Phys.* **101**, 063918 (2007)
[25] R. Ramaneti et al, *Appl. Phys. Lett.,* **91**, 012502 (2007